\documentstyle[11pt,aaspp4]{article}
\input psfig

\begin{document}
\title{Frequency Shifts of the Solar Fundamental Mode}
\author{Andrei Gruzinov }
\affil{Institute for Advanced Study, School of Natural 
Sciences,
Princeton, NJ 08540}

\begin{abstract}

The observed frequency shifts of the high-degree solar fundamental mode are explained using a simple geometrical optics approximation. The predicted fractional frequency shift is $-0.4375<V^2>/v_g^2$, where $<V^2>$ is the mean squared velocity of the convection flow at the surface of the Sun, $v_g$ is the group velocity of the mode. The data converge to the predicted behavior at very high degrees, with $<V^2>^{1/2}=1.3~{\rm km/s}$.

\end{abstract}

\section{Frequency Shift of the High-Degree Solar Fundamental Mode}
The observed frequencies  of the high-degree solar fundamental mode (Duvall et al 1998) are well below the theoretical value $\sqrt{gk}$, where $g$ is the gravitational acceleration, $k=(l+0.5)/R_{\odot }$ is the wavenumber, and $l$ is the degree of the mode. The theoretical frequency is only approximately equal to $\sqrt{gk}$, but the accuracy of this approximation is better than $0.3\%$ for $1000<l<2000$ (Gough 1993).  The measured fractional frequency shift relative to $\sqrt{gk}$ is $\sim -3\%$. The frequencies are measured with an $\sim 0.1\%$ accuracy.  Here we give a quantitative explanation of the measured frequency shift. Our approach - geometrical optics of the surface mode in a random horizontal flow - is much simpler than other theories (Duvall et al 1998 and references therein). A clear general discussion of the effects of convection on the solar oscillation frequencies is given by Rosenthal (1997). Here we just describe our model and show that it matches the data.

We approximate a high-degree solar fundamental mode by a surface (two dimensional) phonon. We represent the effects of convection by a random potential flow on the surface of the Sun. These are reasonable approximations at very high degrees, when the fundamental mode wavelength becomes smaller than the characteristic wavelength of the horizontal flow, that is for $l\gtrsim 1000$. For our purposes, the flow might be taken to be stationary. The phonon propagating through the flow spends more time going against the flow than with the flow. This reduces its effective speed, meaning that the frequency shift is negative. On dimensional grounds, the fractional frequency shift should be $\sim -<V^2>/v_g^2$, where $<V^2>$ is the mean squared velocity of the flow, $v_g$ is the group velocity of the mode. An exact calculation (Appendix A, exact in the leading order in $V$) gives 
\begin{equation}
{\delta \omega \over \omega}=-{7\over 16}{<V^2>\over v_g^2}.
\end{equation}

Since $v_g\propto l^{-1/2}$, we have a nontrivial prediction: {\it at very high degrees, the fractional frequency shift of the fundamental mode should be proportional to $l$}. The high degree data do show this behavior. In Fig.1 we plot the root mean squared velocity of the flow $<V^2>^{1/2}$ calculated from eq. (1), using the frequency shifts calculated from Table 1 of Duvall et al (1998). The inferred root mean squared velocity is 
\begin{equation}
<V^2>^{1/2}=1.3~{\rm km/s}.
\end{equation}

\section{Summary}
The observed frequency shifts of the solar fundamental mode are proportional to $l$ at very high $l$. This behavior is predicted by geometrical optics of a surface mode in a random horizontal flow. The inferred  root mean squared velocity is $<V^2>^{1/2}=1.3~{\rm km/s}$.

\acknowledgements
I thank Dr. Duvall for making the data available prior to publication. I thank Dr. Bachall and Dr. Basu for very useful comments. This work was supported by NSF PHY-9513835.

\appendix 

\section{Phonon in a Random Potential Flow}
We calculate the frequency shift for a two dimensional phonon of dispersion $\Omega (k)$ in a stationary, potential, two dimensional, statistically-homogeneous and isotropic random flow. Our calculation is exact in the leading order in the flow velocity, which is $<V^2>$. 

In this order, the random flow is a collection of statistically independent randomly directed one dimensional flows, say Fourier harmonics of the flow. We will therefore calculate the frequency shift for a one dimensional flow $V(x)\hat{x}$. After averaging the result over the directions of the phonon, we recover a correct answer to the original problem.

The phonon propagation is described by energy and momentum conservation, $k_y={\rm const}$, and 
\begin{equation}
\omega =\Omega(k)+V(x)k_x={\rm const}.
\end{equation}
Define $k_0$, $k_{x0}$, and $\delta k_x$ by $\Omega(k_0)=\omega$, $k_0^2=k_{x0}^2+k_y^2$, and $\delta k_x=k_x-k_{x0}$. From (A1), up to second order in $V$,
\begin{equation}
{k_{x0}\over k_0}{\delta k_x\over k_0}=-{k_{x0}\over k_0}{V\over v_g}+\left( 1-{1\over 2}{k_y^2\over k_0^2}-{1\over 2}{k_0v_g'\over v_g}{k_{x0}^2\over k_0^2}\right) {V^2\over v_g^2},
\end{equation}
where $v_g=\Omega '(k_0)$, $v_g'=\Omega ''(k_0)$. From (A2), taking a spatial (not a time) average, 
\begin{equation}
{k_{x0}\over k_0}{<\delta k_x>\over k_0}=\left( 1-{1\over 2}{k_y^2\over k_0^2}-{1\over 2}{k_0v_g'\over v_g}{k_{x0}^2\over k_0^2}\right) {<V^2>\over v_g^2}.
\end{equation}

The frequency shift is 
\begin{equation}
\delta \omega =-v_g<\delta k>=-v_gk_0{k_{x0}\over k_0}{<\delta k_x>\over k_0},
\end{equation}
and from (A3)
\begin{equation}
{\delta \omega \over \omega }=-{v_g\over v_p}\left( 1-{1\over 2}{k_y^2\over k^2}-{1\over 2}{kv_g'\over v_g}{k_x^2\over k^2}\right) {<V^2>\over v_g^2},
\end{equation}
where $v_p=\Omega /k$. Averaging over directions of ${\bf k}$ gives the answer
\begin{equation}
{\delta \omega \over \omega }=-{v_g\over v_p}\left( {3\over 4}-{1\over 4}{kv_g'\over v_g}\right) {<V^2>\over v_g^2}.
\end{equation}
For the f-mode, $\Omega \propto k^{1/2}$, and (A6) gives eq. (1).

\begin{figure}[htb]
\psfig{figure=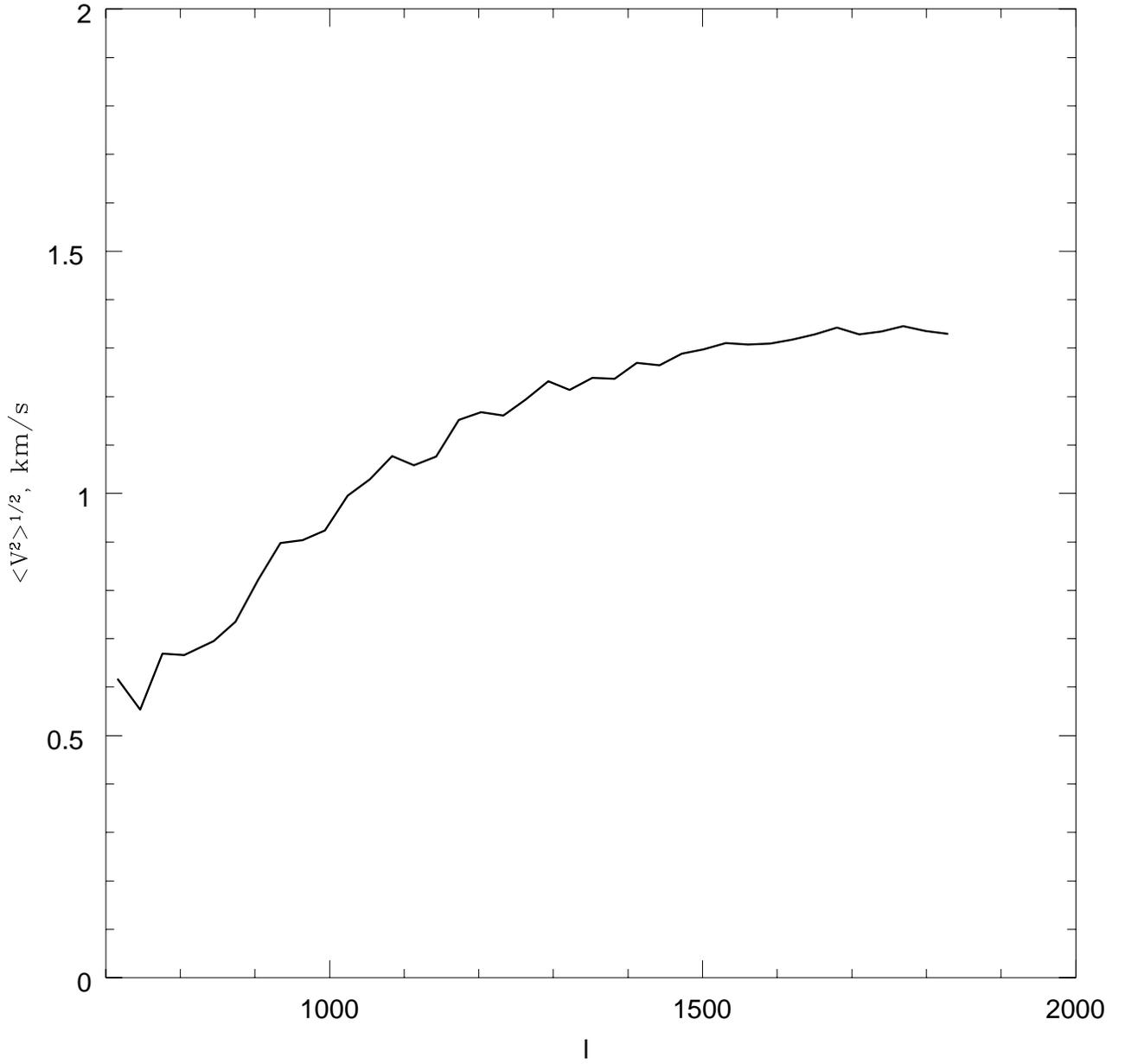,width=7in}
\caption{Calculated root mean squared horizontal velocity $<V^2>^{1/2}$, in km/s, as a function of degree $l$.}
\end{figure}

\end{document}